\documentclass[english,twocolumn,floats,showpacs,amsmath,amssymb,prb]{revtex4}
\usepackage[T1]{fontenc}
\usepackage[latin9]{inputenc}
\usepackage{textcomp}
\usepackage{amsmath}
\usepackage{amssymb}
\usepackage{graphicx}

\makeatletter

\newcommand{\lyxmathsym}[1]{\ifmmode\begingroup\def\b@ld{bold}
  \text{\ifx\math@version\b@ld\bfseries\fi#1}\endgroup\else#1\fi}

\providecommand{\tabularnewline}{\\}

\@ifundefined{textcolor}{}
{%
 \definecolor{BLACK}{gray}{0}
 \definecolor{WHITE}{gray}{1}
 \definecolor{RED}{rgb}{1,0,0}
 \definecolor{GREEN}{rgb}{0,1,0}
 \definecolor{BLUE}{rgb}{0,0,1}
 \definecolor{CYAN}{cmyk}{1,0,0,0}
 \definecolor{MAGENTA}{cmyk}{0,1,0,0}
 \definecolor{YELLOW}{cmyk}{0,0,1,0}
}

\usepackage{epsfig}

\usepackage{babel}

\makeatother

\usepackage{babel}
\begin{document}

\title{The phase diagram of ice: a quasi-harmonic study based on a flexible
water model}

\author{R. Ramírez$^{a)}$, N. Neuerburg, and C. P. Herrero\let\oldthefootnote\thefootnote\global\long\def\thefootnote{{a)}}
 \footnotetext{Electronic mail: ramirez@icmm.csic.es}\let\thefootnote\oldthefootnote }

\affiliation{Instituto de Ciencia de Materiales de Madrid (ICMM), Consejo Superior
de Investigaciones Científicas (CSIC), Campus de Cantoblanco, 28049
Madrid, Spain }

\date{{\today}}
\begin{abstract}
The phase diagram of ice is studied by a quasi-harmonic approximation.
The free energy of all experimentally known ice phases has been calculated
with the flexible q-TIP4P/F model of water. The only exception is
the high pressure ice X, in which the presence of symmetric O$-$H$-$O
bonds prevents its modeling with this empirical interatomic potential.
The simplicity of our approach allows us to study ice phases at state
points of the $T-P$ plane that have been omitted in previous simulations
using free energy methods based on thermodynamic integration. The
effect in the phase diagram of averaging the proton disorder that
appears in several ice phases has been studied. It is found particularly
relevant for ice III, at least for cell sizes typically used in phase
coexistence simulations. New insight into the capability of the employed
water model to describe the coexistence of ice phases is presented.
We find that the H-ordered ices IX and XIV, as well as the H-disordered
ice XII, are particularly stable for this water model. This fact disagrees
with experimental data. The unexpected large stability of ice IX is
a property related to the TIP4P-character of the water model. Only
after omission of these three stable ice phases, the calculated phase
diagram becomes in reasonable qualitative agreement to the experimental
one in the $T-P$ region corresponding to ice Ih, II, III, V, and
VI. The calculation of the phase diagram in the quantum and classical
limits shows that the most important quantum effect is the stabilization
of ice II due to its lower zero-point energy when compared to that
one of ices Ih, III, and V.
\end{abstract}


\maketitle

\section{Introduction\label{sec:intro}}

Sixteen different crystalline ice phases have been identified so far
in the phase diagram of water.\cite{dunaeva10} In all phases, except
ice X, the water molecules are part of a network connected by H-bonds.
In most ice lattices there appears a unique H-bond network that fills
the whole volume. However four phases (ices VII, VIII, VI, and XV)
are made of two identical and independent networks that interpenetrate
one into another. Within a H-bond network, each oxygen atom is coordinated
to four oxygen neighbors in a more or less distorted tetrahedral arrangement.
The protons are distributed according to the Bernal-Fowler ice rules.
They state that in a network there must be one and only one proton
between two adjacent oxygen atoms and that each oxygen is part of
two OH covalent bonds characteristic of the water molecule. \cite{bernal33}
These rules are compatible with either ordered or disordered spatial
distributions of H atoms. In fact order-disorder transitions have
been observed for most pairs of ice phases (Ih-XI, III-IX, V-XIII,
VI-XV, VII-VIII, XII-XIV). Only for H-ordered ice II and H-disordered
ices Ic and IV, the other member of the corresponding pair has not
been yet experimentally found.\cite{salzmann11} 

A comprehensive review of the calculation of free energies of water
phases with the thermodynamic integration (TI) method can be found
in Ref. \onlinecite{vega09}. The classical phase diagram of water,
simulated with the rigid TIP4P/2005 model, shows a reasonable qualitative
agreement to the experimental one, in particular in the complex region
of stability of ices Ih, II, III, V, and VI.\cite{abascal05} The
coexistence of these ice phases has been also studied by quantum path
integral simulations with the rigid TIP4PQ/2005 model.\cite{mcbride12}
The phase diagram of ice Ih, II, and III was additionally investigated
using a flexible water model (q-TIP4P/F) in the classical limit.\cite{habershon11}
Singer and Knight have analyzed the order-disorder transition in ices
Ih-XI, III-IX, V-XIII, VI-XV, VII-VIII, and XII-XIV by the calculation
of the small energy differences between the innumerable H-bond configurations
possible in a large simulation cell.\cite{singer12} Since the lattice
energy is a scalar, it can be related to topological properties of
the H-bond configurations that are invariant to the symmetry operations
of the lattice. This link between H-bond topology and energetics is
used to extrapolate from electronic calculations on small unit cells
to larger cells that approximate the thermodynamic limit. Thus accurate
results for the order-disorder transitions in ice are obtained just
by focusing on the dependence of the lattice energy with the H-bond
configurations. The vibrational energy was assumed to play a secondary
role in these transitions. H-bond order-disorder transitions are understood
as discontinuous changes in the H-bond topologies sampled by the system,
while the oxygen lattice changes minimally. Note that for phase transitions
other than order-disorder ones, the change in the oxygen lattice is
drastic. Therefore for such transitions the vibrational free energy
is expected to play a significant role.

The quasi-harmonic (QH) approximation (QHA) allows us to compute the
free energy of a solid as an analytic function of the volume and the
temperature for a given interatomic potential.\cite{srivastava} The
prediction of the volume, enthalpy, kinetic energy, and heat capacity,
of ice Ih, II, and III by the QHA has been compared to results of
quantum path integral molecular dynamics (PIMD) simulations using
the q-TIP4P/F model. A remarkable overall agreement was found in temperature
($T$) and pressure ($P$) ranges up to 300 K and 1 GPa, respectively.\cite{ramirez12}
Moreover, the QHA offers a simple alternative to TI methods to study
the phase diagram of solid phases. 

The phase boundary between ice VII and VIII has been studied by a
QHA in a 16-molecule supercell with \textit{ab initio} density-functional
theory (DFT) calculations of total energies and phonon frequencies.\cite{umemoto10}
The calculation shows that the coexistence line in the $P-T$ diagram
has negative Clapeyron slope and a noticeable isotope effect, both
facts in good agreement to experimental data. The phase diagram of
ices Ih, II, and III has been recently calculated by a QHA.\cite{ramirez12b}
The studied models were based on both flexible (q-TIP4P/F) and rigid
(TIP4P/2005 and TIP4PQ/2005) descriptions of the water molecule. The
QHA was able to reproduce, for each of the studied models, the available
coexistence lines Ih-II, II-III, and Ih-III of the phase diagrams
derived by TI methods. Moreover, the simplicity of the QHA allowed
to uncover new information by considering conditions that had not
been treated in previous TI simulations. In particular, for the typical
cell sizes used in phase coexistence simulations, the averaging over
the proton disorder of ice III was an essential step to obtain a converged
phase diagram. Thus, the common procedure of using only one randomly
selected ice III structure makes the calculated phase diagram affected
by an uncontrolled factor, that can be highly significant for the
final result.\cite{ramirez12b} 

The purpose of the present paper is to derive the phase diagram of
all experimentally known ice phases of ice (except ice X) using the
QHA in combination with the flexible q-TIP4P/F water model.\cite{habershon09}
The layout of the manuscript is as follows. A summary of the employed
computational conditions is presented in Sec. \ref{sec:Computational-conditions}.
The generation of the ice structures is introduced in Sec. \ref{sec:ice structures}.
The effect of H-disorder averaging in the lattice energy is discussed
in Sec. \ref{sec:Disorder-averaging}. The calculated phase diagram
of ice is compared to the experimental one in Sec. \ref{sec:Phase-diagram}.
The pressure dependence of the free energy of several ice phases is
presented in Sec. \ref{sec:QH-free-energies}. A comparison of classical
and quantum phase diagrams is given in Sec. \ref{sec:Quantum-vs.-classical}.
The paper closes with the conclusions.

\section{Computational conditions\label{sec:Computational-conditions}}

The employed QHA has been introduced in Refs. \onlinecite{ramirez12}
and \onlinecite{ramirez12b}. We present here a brief summary. The
Helmholtz free energy of an ice phase with $N$ water molecules in
a cell of volume $V$ and at temperature $T$ is given by 

\begin{equation}
F(V,T)=U_{S}(V)+F_{v}(V,T)-TS_{H}+\triangle U_{ave}\;,\label{eq:f_v_t}
\end{equation}
where $U_{S}(V)$ is the static zero-temperature classical energy,
i.e., the minimum of the potential energy when the volume of the cell
is $V$. $F_{v}(V,T)$ is the vibrational contribution to $F$. In
the quantum limit it is given by

\begin{equation}
F_{v}(V,T)=\sum_{k}\left(\frac{\hbar\omega_{k}}{2}+\frac{1}{\beta}\ln\left[1-\exp\left(-\beta\hbar\omega_{k}\right)\right]\right)\;.\label{eq:fv_q}
\end{equation}
Here $\beta$ is the inverse temperature: $1/k_{B}T$. $\omega_{k}$
are the wavenumbers of the harmonic lattice vibrations for the volume
$V$, with $k$ combining the phonon branch index and the wave vector
within the Brillouin zone. The anharmonicity of the interatomic potential
enters in the QHA only through the volume dependence of $\omega_{k}$.
In the classical limit the vibrational contribution amounts to

\begin{equation}
F_{v,cla}(V,T)=\sum_{k}\frac{1}{\beta}\ln\left(\beta\hbar\omega_{k}\right)\;.\label{eq: fv_c}
\end{equation}
The entropy $S_{H}$ and the energy $\triangle U_{ave}$ are related
to the disorder of hydrogen and they vanish for the ordered ice phases
(i.e., ices XI, II, IX, XIII, XV, VIII, and XIV) . $S_{H}$ was estimated
by Pauling for fully disordered phases as \citep{pauling35}

\begin{equation}
S_{H}=Nk_{B}\ln\frac{3}{2}\;.\label{eq:entropy}
\end{equation}
A comparison of the Pauling estimate to accurate numerical determinations
has been recently presented for several ice phases.\cite{herrero13}
$\triangle U_{ave}$ is a constant energy that depends on the average
of the lattice energy over the proton disorder of the ice phase (see
below). The Gibbs free energy, $G(T,P),$ is obtained by seeking for
the volume, $V_{min}$, that minimizes the function $F(V,T)+PV$,
as

\begin{equation}
G(T,P)=F(V_{min},T)+PV_{min}\;.\label{eq: G}
\end{equation}
The implementation of the QHA for an ice phase follows these steps:\cite{ramirez12,ramirez12b} 

$i)$ Find the reference cell that minimizes the static energy $U_{S}$.
This minimization implies optimization of both cell shape and atomic
positions. The resulting volume is $V_{ref}$ and the corresponding
static energy $U_{S,ref}$.

$ii)$ Only for H-disordered phases: generate a random set of structures
with different H-configurations and calculate the constant energy
$\triangle U_{ave}$ as

\begin{equation}
\triangle U{}_{ave}=\overline{U}_{S,ref}-U_{S,ref}\;,\label{eq: U_ave}
\end{equation}
where $\overline{U}_{S,ref}$ is the average of the static lattice
energy for the generated set of H-isomers, while $U_{S,ref}$ is the
lattice energy of the reference cell considered in step $i)$. The
number of random H-isomers is set so large that the estimated error
of the mean value $\overline{U}_{S,ref}$ is lower than 0.02 kJ/mol.
It is also sensible to take as reference cell in step $i)$ the structure
whose lattice energy $U_{S,ref}$ is closest to the average $\overline{U}_{S,ref}$
.

$iii)$ Select a grid of 50 volumes in a range of interest $\left[V_{min},V_{max}\right]$.
The ice cell with volume $V_{i}$ is set by isotropic scaling of the
reference cell. Subsequently, each ice cell is held fixed while minimizing
the static energy $U_{S}(V_{i})$ with respect to the atomic positions.
The crystal phonons, $\omega_{k}(V_{i})$, are obtained after the
minimization. 

$iv)$ Calculate the function $F(V_{i},T)$ by Eq. (\ref{eq:f_v_t}).
The minimum of $F(V_{i},T)$ as a function of $V$ is determined by
a fit to a 5th degree polynomial in $V$. 

The phase diagram of ice is derived by a brute force method, i.e.,
given a state point $(T,P)$ one calculates the Gibbs free energy
of all ice phases. The stable phase is selected as the one with the
lowest value of $G$. 

The crystal phonon calculation has been performed by the small-displacement
method.\citep{kresse95,alfe01} For the flexible water model the atomic
displacement employed in this work is $\delta x=10^{-6}$ $\textrm{\AA}$
along each Cartesian direction. We have used a $\Gamma$ sampling
(\textbf{$\mathbf{k}=\mathbf{0}$}) of the crystal phonons, which
is a reasonable approximation for the sizes of the employed supercells.
The Lennard-Jones interaction between oxygen centers was truncated
at a distance of $r_{c}=8.5\textrm{ \AA}$, and standard long-range
corrections for both potential energy and pressure were computed assuming
that the pair-correlation function is unity for $r>r_{c}$.\cite{johnson93}
Long-range electrostatic potential and forces were calculated with
the Ewald method. 

The assumption of isotropic scaling of the reference cell made in
step $iii$ was checked for ice II in Ref. \onlinecite{ramirez12b}.
By relaxing this constraint the QHA free energy of ice II changes
slightly, by about 0.01 kJ/mol, having a small effect in the phase
diagram.\cite{ramirez12b}

\begin{table*}
\caption{The space group of the studied phases is shown with the
reference
used to generate the ice supercell. H-disordered phases (except ice
IV and ice Ic) are found in a row immediately above the H-ordered
counterpart. The number of water molecules in each supercell is $N$.
The static lattice energy ($U_{S,ref}$) and volume ($V_{ref}$ )
of the optimized supercells were derived with the q-TIP4P/F model.
$\left[V_{min},V_{max}\right]$ is the volume interval studied by
the QHA. The data for ices III and V correspond to both full and
partial
H-disorder. Energy unit is kJ/mol, volumes in
$\textrm{\AA}^{3}/\mathrm{molec}.$}
\label{tab:all_ices}

\vspace{3mm}
\begin{tabular}{lllccccc}
\hline
Space symmetry & Supercell & H-disorder & $N$ & $\quad
U_{S,ref}\quad$ & $\quad$$V_{ref}$$\quad$ & $V_{min}$ & $\quad
V_{max}\quad$\tabularnewline
\hline
Ic ($Fd\overline{3}m$)\cite{konig44} & (3, 3, 3) & yes & 216 &
-62.00 & 31.00 & 20.44 & 35.02\tabularnewline
Ih $(P6_{3}/mmc$)\cite{peterson57} & (4, 3$\sqrt{3},$ 3) & yes & 288
& -61.98 & 30.96 & 29.47 & 35.05\tabularnewline
XI ($Cmc2_{1}$)\cite{leadbetter85} & (4, 3, 3) & no & 288 & -61.95 &
31.03 & 29.48 & 35.05\tabularnewline
XI ($Pna2_{1}$)\cite{hirsch04} & (3, 3, 4) & no & 288 & -62.02 &
30.90 & 29.30 & 34.96\tabularnewline
II ($R\overline{3}$)\cite{kamb71} & (3, 3, 3) & no & 324 & -60.84 &
24.14 & 21.75 & 27.31\tabularnewline
III  ($P4_{1}2_{1}2$)  & (3, 3, 3) & yes (full) & 324 & -60.96 &
24.90 & 23.58 & 28.14\tabularnewline
III ($P4_{1}2_{1}2$) \cite{lobban00} & (3, 3, 3) & yes (partial) &
324 & -60.72 & 25.05 & 24.07 & 28.31\tabularnewline
IX ($P4_{1}2_{1}2$) \cite{laplaca73} & (3, 3, 3) & no & 324 & -61.52
& 24.63 & 23.55 & 27.83\tabularnewline
IV ($R\overline{3}c$) \cite{engelhardt81} & (2, 2, 2) & yes & 128 &
-59.77 & 22.10 & 18.49 & 24.31\tabularnewline
V ($C2/c$)  & (2, 3, 3) & yes (full) & 504 & -60.28 & 22.84 & 20.55
& 25.81\tabularnewline
V ($C2/c$) \cite{lobban00} & (2, 3, 3) & yes (partial) & 504 &
-60.04 & 22.99 & 20.80 & 25.98\tabularnewline
XIII ($P2_{1}/a$)\cite{salzmann06} & (2, 3, 3) & no & 504 & -60.15 &
23.15 & 20.83 & 26.16\tabularnewline
VI $(P4_{2}/nmc$)\cite{kuhs84} & (3, 3, 4) & yes & 360 & -59.58 &
21.44 & 18.14 & 22.51\tabularnewline
XV ($P\overline{1}$)\cite{salzmann09} & (3, 3, 4) & no & 360 &
-59.43 & 21.53 & 18.43 & 22.61\tabularnewline
VII ($Pn\overline{3}m$)\cite{kuhs84} & (6, 6, 6) & yes & 432 &
-53.08 & 19.57 & 14.02 & 20.98\tabularnewline
VIII ($I4_{1}/amd$) \cite{besson94} & (5, 5, 3) & no & 600 & -53.19
& 19.47 & 13.92 & 20.83\tabularnewline
XII ($I\overline{4}2d$)\cite{lobban98} & (2, 2, 4) & yes & 288 &
-60.06 & 21.99 & 18.81 & 24.19\tabularnewline
XIV ($P2_{1}2_{1}2_{1}$) \cite{salzmann06} & (2, 2, 4) & no & 192 &
-60.62 & 21.90 & 18.32 & 24.09\tabularnewline
\hline
\end{tabular}
\end{table*}

\section{Ice structures\label{sec:ice structures}}

Supercells of similar size to those employed in recent simulations\cite{habershon11,mcbride12}
have been used in the QH derivation of the phase diagram. In Table
\ref{tab:all_ices} we summarize the crystallographic references used
in the generation of the ice structures. Supercells are defined by
translation vectors applied along the crystallographic axes of the
lattice. The total number $N$ of water molecules generated in the
supercell is also given. The potential energy ($U_{S,ref}$) and volume
($V_{ref}$) obtained in the minimization of the supercell structures
with the q-TIP4P/F model are presented, together with the volume interval
{[}$V_{min},V_{max}${]} used in the QHA for each phase. The optimized
reference cells and the corresponding fractional coordinates of the
water molecules for each studied ice phase are made available as Supplementary
Material.\cite{supplementary}

The algorithm proposed by Buch \textit{et al.} in Ref. \onlinecite{buch98}
was applied for the random generation of full proton disordered structures
with vanishing cell dipole moment of ices Ih, Ic, III, IV, V, VI,
VII, and XII. The reason for choosing a vanishing cell dipole moment
is that the disordered ice phases are not ferroelectric. In the case
of ice III and ice V the neutron diffraction experiments show the
existence of partial H-disorder, i.e., fractional H-occupancies different
from $0.5$.\cite{lobban00} The Buch's algorithm has been then slightly
modified for the generation of random structures with partial H-disorder.\cite{macdowell04}
For these phases the proton disorder entropy is somewhat lower than
the Pauling result, $S_{H}$. We have employed the estimations of
$0.9S_{H}$ and $0.94S_{H}$ for ice III and V, respectively.\cite{macdowell04} 

In the generation of the crystal structures the following particularities
were considered. For ice Ih the reference supercell was orthorhombic
with parameters $(4\mathbf{a},3\sqrt{3}\mathbf{b},3\mathbf{c})$,
with $(\mathbf{a},\mathbf{b},\mathbf{c})$ being the standard hexagonal
lattice vectors of ice Ih.\cite{hayward87} For ice XI, the H-ordered
form of ice Ih, we have generated two different structures with crystal
symmetry $Cmc2_{1}$ and $Pna2_{1}$. The former corresponds to the
experimental phase ice XI.\cite{leadbetter85} The latter is associated
to the global energy minimum predicted by TIP4P-like models, that
is not in accord with experiment.\cite{buch98} For ice V the orthorhombic
cell setting $(\mathbf{a},\mathbf{b},\mathbf{c})$ used in Ref. \onlinecite{lobban00}
corresponds to the space group symbol $A2/a$, but was changed here
to a more standard setting $(\mathbf{c},\mathbf{-b},\mathbf{a})$
with space group symbol $C2/c$.\cite{ITC83} In the case of the high
pressure phases ice VII and VIII, the energy minimization of a flexible
supercell did not lead to a stable crystal lattice at zero pressure.
To overcome this difficulty the form of the supercell was constrained
to that one obtained by classical $NPT$ simulations of ice VII and
VIII using a flexible cell at $P=$2 GPa and $T=$50 K, i.e., at conditions
where instability problems are fully absent. Then subsequent energy
minimizations of ices VII and VIII maintaining the rigid form of the
ice cell do always lead to stable crystal structures even at volumes
corresponding to small negative pressures.

\begin{table}
\caption{Ice densities derived by PIMD simulations of ice phases using 
the q-TIP4P/F model are compared to the corresponding QHA, as well as
to experimental data. Simulation results for ices III and V correspond
to cells with full H-disorder. The employed H-isomer of ice III had
a static lattice energy of -60.86 kJ/mol.\cite{ramirez12b} Density
unit is g cm$\mathrm{}^{-3}$.}
\label{tab:density}

\vspace{3mm}
\begin{tabular}{ccccccc}
\hline
Phase & $T$ (K) & $P$(GPa) & $\rho$(PIMD) & $\rho$ (QHA) & $\rho$
(exp.) & Ref.\tabularnewline
\hline
Ih & 250 & 0.0 & 0.925 & 0.917 & 0.920 &
\cite{rottger94}\tabularnewline
II & 123 & 0.0 & 1.190 & 1.191 & 1.190 &
\cite{fortes05}\tabularnewline
III (full) & 250 & 0.28 & 1.168 & 1.177 & 1.165 &
\cite{londono93}\tabularnewline
IV & 110 & 0.0 & 1.290 & 1.296 & 1.272 &
\cite{engelhardt81}\tabularnewline
V (full) & 237.7 & 0.53 & 1.269 & 1.272 & 1.271 &
\cite{gagnon90}\tabularnewline
VI & 225 & 1.1 & 1.397 & 1.382 & 1.373 &
\cite{kuhs84}\tabularnewline
VII & 300 & 10 & 1.783 & 1.785 & 1.880 &
\cite{hemley87}\tabularnewline
VIII & 77 & 2.4 & 1.590 & 1.592 & 1.628 &
\cite{kuhs84}\tabularnewline
IX & 165 & 0.28 & 1.187 & 1.191 & 1.194 &
\cite{londono93}\tabularnewline
XI & 77 & 0.0 & 0.931 & 0.930 & 0.934 & \cite{line96}\tabularnewline
XII & 260 & 0.5 & 1.301 & 1.299 & 1.292 &
\cite{lobban98}\tabularnewline
XIII & 80 & 0.0 & 1.241 & 1.242 & 1.244 &
\cite{salzmann06}\tabularnewline
XIV & 250 & 0.28 & 1.308 & 1.311 & 1.332 &
\cite{salzmann06}\tabularnewline
XV & 80 & 0.0 & 1.329 & 1.336 & 1.326 &
\cite{salzmann09}\tabularnewline
\hline
\end{tabular}
\end{table}

The validity of the QHA in ice is restricted by the presence of anharmonic
effects not included in the approximation. Such effects are expected
to increase at high temperature. A direct check of the QHA is the
comparison to numerical simulations that fully consider the anharmonicity
of the interatomic interactions. PIMD results of the density of ice
phases for a number of state points are compared to the corresponding
QHA as well as to available experimental data in Table~\ref{tab:density}.
The studied state points appear in a temperature range between 77
and 300 K. We find a reasonable agreement between PIMD and QHA densities
even at high temperature. Similar temperature behavior was reported
for the volume, enthalpy, and heat capacity of ices Ih, II, and III
in Ref. \onlinecite{ramirez12b} up to 300 K. Note that the thermal
energy at 300 K corresponds to a wavenumber ($k_{B}T/\hbar)$ of about
200 cm$\mathrm{^{-1}}$, so that at 300 K most ice phonons remain
in their vibrational ground state. In particular, those related to
the molecular stretching and bending modes, as well as the H-bond
librations.\cite{ramirez12} The comparison between calculated and
experimental ice densities in Table~\ref{tab:density} displays a 
satisfactory
overall agreement. The largest deviation is found for the high-pressure
ice VII, where the calculated q-TIP4P/F density is about 5\% lower
than the experimental one. Such error has been previously reported
in ice VII by classical and quantum Monte Carlo (MC) simulations using
TIP4P-like models.\cite{aragones2009,mcbride09}

\section{Disorder averaging\label{sec:Disorder-averaging}}

An interesting practical question is the importance of proton disorder
in the stability of H-disordered cells. Two opposite strategies have
been used to address this point. The first one is to use a cell so
small that an explicit calculation of the internal energy of all existing
H-isomers is possible. The lattice symmetry may be exploited by considering
only symmetry inequivalent H-isomers with the corresponding multiplicity.\cite{umemoto10}
This approach leads to an exact average over proton disorder, but
at the cost of introducing an unspecified finite size effect as a
consequence of the small unit cell. 

The second strategy is to define a large supercell and calculate the
internal energy of a \textit{single H-isomer} of the disordered phase.
The average over proton disorder is then introduced \textit{ad hoc
}by adding the proton disorder entropy $S_{H}$, as in Eq. \ref{eq:f_v_t}.
An implicit assumption here is that the supercell is so large that
both the lattice energy and the vibrational free energy of the single
H-isomer have already converged with respect to any change in the
proton configuration. Such a change is understood to be compatible
with the Bernal-Fowler rules and with the corresponding (full or partial)
proton disorder. This procedure has been adopted in most TI studies
of the phase diagram of ice.\cite{sanz04,abascal05,habershon11,mcbride12}
In addition we have already commented in the Introduction the approach
by Singer and Knight, extrapolating the results of small unit cells
to larger ones to study order-disorder transitions. \cite{singer12}

Note that the assumption of the convergence of the lattice energy
with the proton disorder in a large supercell is only correct in the
thermodynamic limit, as the relative fluctuation of thermodynamic
quantities is expected to decrease as $1/\sqrt{N}$. For finite $N,$
the Buch's algorithm will produce a set of H-isomers with different
lattice energies and therefore also different statistical weights.
Only in the thermodynamic limit will the Buch's algorithm produce
proton configurations with the same statistical weight, as it was
assumed by Pauling in his estimation of the residual entropy of ice
in Eq. (\ref{eq:entropy}). In this respect, Eq. (\ref{eq: U_ave})
can be understood as a finite size correction for the estimation of
the lattice energy associated to the thermodynamic limit.\cite{ramirez12b}

The simplicity of the QHA renders possible to check the convergence
of the lattice energy, for a given cell size, with respect to the
proton disorder. In our previous QH study of ice Ih and III, the convergence
of the static lattice energy, $U_{S,ref}$, was studied for a small
set of six random H-isomers in a cell with 324 molecules.\cite{ramirez12b}
In the following we present a more accurate account of the convergence
of $U_{S,ref}$ for ice III.

\subsection{Ice III}

\begin{figure}
\vspace{-1.8cm}
\includegraphics[width= 9cm]{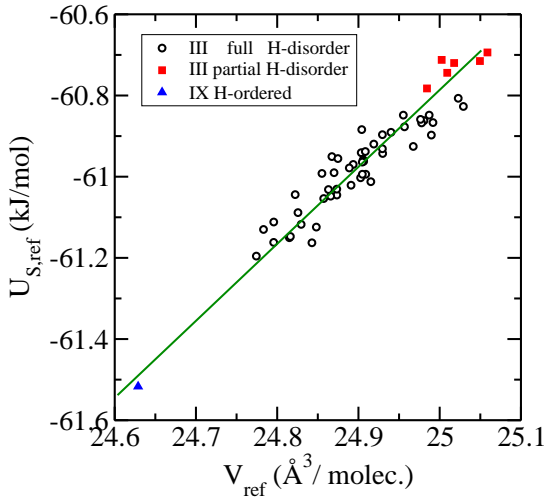}
\vspace{-0.9cm}
\caption{Lattice energy and volume of a set of H-isomers of ice III
generated
randomly according to the Bernal-Fowler ice rules. The H-isomers display
either full (open circles) or partial (closed squares) H-disorder.
The results were derived with the q-TIP4P/F model for a supercell
with 324 molecules. The close triangle shows the result for ice IX,
the H-ordered counterpart of ice III. The line is a guide to the eye. }
\label{fig: U_V_ref}
\end{figure}

In Fig. \ref{fig: U_V_ref} we have represented the results of $U_{S,ref}$
for a set of 50 random H-isomers of ice III. The supercell contains
324 water molecules with full H-disorder. The static lattice energy
$U_{S,ref}$ is plotted as a function of the corresponding cell volume,
$V_{ref}$. For comparison, we show also the data obtained when ice
III has partial H-disorder, and the value for ice IX, the H-ordered
counterpart of ice III. The minimized lattice energy, $U_{S,ref}$,
and volume, $V_{ref}$, are related in a nearly linear way. We note
that all isomers having partial H-disorder display \textit{larger}
static energy than those with full H-disorder. 

An important result of Fig. \ref{fig: U_V_ref} is that for the employed
supercell with 324 molecules the dispersion of $U_{S,ref}$ is rather
large ($\sim0.4$ kJ/mol). We have chosen a threshold of 0.05 kJ/mol
as criterion to qualify if a given energetic difference can be considered
as significant in the calculation of the phase diagram. Then, following
this criterion, the spreading of the lattice energy of ice III may
appreciably affect the phase diagram whenever it is calculated with
a \textit{single random} H-isomer of ice III. As a remedy to this
uncertainty, the average term $\triangle U_{ave}$ was introduced
in Eq. \ref{eq:f_v_t} to improve the convergence of the internal
energy of ice with respect to the proton disorder. \cite{ramirez12b}

\begin{figure}
\vspace{0.2cm}
\hspace{-4.5mm}
\includegraphics[width= 5cm]{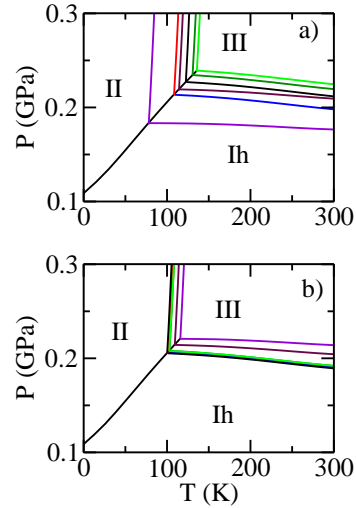}
\vspace{0.3cm}
\caption{$a)$ QHA phase diagram of ice Ih, II, and III. The coexistence
lines
were calculated for six random H-isomers of ice III with full
H-disorder.
The free energy of ice III does not include the disorder averaging
term $\triangle U_{ave}$ of Eq. (\ref{eq:f_v_t}). $b)$ Same diagram
after including the disorder averaging term $\triangle U_{ave}$ in
the free energy of ice III. Note the decrease in the dispersion of
the coexistence lines. All results derived with the q-TIP4P/F model
in the quantum limit.}
\label{fig: I_II_III}
\end{figure}

The effectiveness of this averaging procedure is illustrated in Fig.
\ref{fig: I_II_III}. It shows the phase diagram of ice Ih, II, and
III calculated for six random H-isomers of ice III with the q-TIP4P/F
model. The lattice energy $U_{S,ref}$ of these H-isomers scatters
in an interval of about 0.3 kJ/mol. In Fig. \ref{fig: I_II_III}a
the free energy of ice III was calculated \textit{without} the H-disorder
averaging term ($\triangle U_{ave}$) of Eq. \ref{eq:f_v_t}. These
results are identical to those presented in Fig.4 of 
Ref.~\onlinecite{ramirez12b}.
Coexistence lines of different ice III structures are clearly separated.
The region of stability of ices Ih, II, and III appears strongly dependent
on the H-disorder. The phase diagram obtained after considering the
term $\triangle U_{ave}$ in Eq. \ref{eq:f_v_t} is shown in Fig.
\ref{fig: I_II_III}b. The spreading of the coexistence lines is now
appreciably reduced. The remaining dispersion is related to the vibrational
free energy, that is also affected by the disorder of protons in the
employed supercell. However, this effect of H-disorder in the vibrational
energy is comparatively less important than in the lattice energy.
A similar conclusion has been presented in the analysis of order-disorder
transitions in Ref. \onlinecite{singer12}. 

\begin{table}
\vspace{-3mm}
\caption{Result of the averaging of the lattice energy, $U_{S,ref},$
of H-disordered
ices using the q-TIP4P/F model. For each phase we show the number
of molecules in the supercell ($N$), the number of random H-isomers
in the average ($N_{ave}$), the mean value of the lattice energy,
$\overline{U}_{S,ref}$, and its standard deviation,
$\sigma(U_{S,ref})$.
Results for ices III and V are shown for both full and partial
H-disorder.
The last two columns in units of kJ/mol.}
\label{tab:mean_U_ref}
\vspace{3mm}
\begin{tabular}{ccccc}
\hline
Ice & $\quad N\quad$ & $\quad N_{ave}$ &
$\quad\overline{U}_{S,ref\quad}$ & $\sigma(U_{S,ref})$\tabularnewline
\hline
Ih  & 288 & 6 & -61.98 & 0.00\tabularnewline
Ic  & 216 & 9 & -62.00 & 0.00\tabularnewline
III (full) & 324 & 50 & -60.98 & 0.10\tabularnewline
III (partial) & 324 & 6 & -60.73 & 0.03\tabularnewline
IV  & 128 & 9 & -59.77 & 0.02\tabularnewline
V (full) & 504 & 10 & -60.27 & 0.03\tabularnewline
V (partial) & 504 & 6 & -60.03 & 0.01\tabularnewline
VI  & 360 & 10 & -59.57 & 0.02\tabularnewline
VII  & 432 & 6 & -53.08 & 0.01\tabularnewline
XII  & 288 & 9 & -60.07 & 0.04\tabularnewline
\hline
\end{tabular}
\end{table}

\subsection{Other disordered phases}

The mean static lattice energy, $\overline{U}_{S,ref}$, and its standard
deviation, $\sigma(U_{S,ref})$, was calculated by sampling a set
of random H-isomers for all H-disordered phases (Ih, Ic, III, IV,
V, VI, VII, and XII). The results are summarized in 
Table~\ref{tab:mean_U_ref}.
A large value of the standard deviation, $\sigma$, implies that H-disorder
strongly affects the value of the static energy $U_{S,ref}$, of the
supercell and therefore also the stability of the ice phase. The largest
value of $\sigma$ corresponds to ice III with full H-disorder ($\sigma=$0.1
kJ/mol ). Accordingly the static energy, $U_{S,ref}$, of a single
random H-isomer of ice III can be found in an energy window of about
$4\sigma\sim0.4$ kJ/mol, as shown in Fig. \ref{fig: U_V_ref}.

The standard deviations $\sigma(U_{S,ref})$ given in 
Table~\ref{tab:mean_U_ref}
decrease along the series of ices: III (full disorder) > XII > V (full
disorder) > III (partial disorder) > IV > VI. For these phases, the
\textit{no} \textit{consideration} of H-disorder averaging may introduce
arbitrary shifts in the lattice energy larger than 0.05 kJ/mol, at
least for the supercell sizes employed here. For ices Ih, Ic, and
VII the energetic effect of H-disorder is smaller than this threshold
so that it can be safely neglected for the studied supercells. 

For ices III and ice V, the mean lattice energy, $\overline{U}_{S,ref}$,
is significantly larger ($\sim$ 0.24 kJ/mol) in the case of partial
than in the case of full H-disorder. This behavior is in contradiction
to the experimental finding that ice III and V display both partial
H-disorder.\cite{lobban00} This unphysical result is in line with
the reported limitations of the effective potentials to reproduce
the energetics of the H-bond rearrangement in the ice phases.\cite{buch98,singer12} 

Our analysis on the disorder averaging of the lattice energy of ice
has omitted several factors that might be relevant. The consideration
of reference cells with non-zero dipole moment should increase the
standard deviation, $\sigma(U_{S,ref})$, and affect also the mean
static lattice energy of the cell. This behavior has been demonstrated
in the classical MC simulation of the dielectric constant of ice using
several water models.\cite{aragones2011} Another factor is the fractional
occupation of H-sites in the ice structures with partial H-disorder
(ices III and V), that may depend on the employed water model. This
was shown in Ref. \onlinecite{aragones2011} where the fractional
occupancies (H$_{\alpha}$, H$_{\beta}$) of ice III, experimentally
found as (0.35, 0.5),\cite{lobban00} change to (0.5, 0.25) by using
the TIP4P/2005 model.\cite{aragones2011} In our treatment of partial
disorder in ice III and ice V we have considered only fractional occupancies
derived from the experimental data.

\section{Quasi-harmonic phase diagram \label{sec:Phase-diagram}}

\begin{figure}
\vspace{-1.8cm}
\includegraphics[width= 9cm]{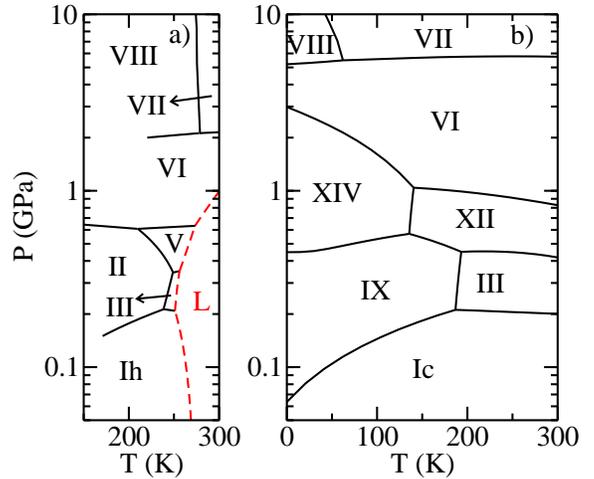}
\vspace{-0.9cm}
\caption{Phase diagram of ice. $a)$ experimental result from Ref.
\onlinecite{dunaeva10}.
The broken line is the boundary between ice and liquid (L) water.
Full lines denote coexistence conditions between ice phases. $b)$
QHA result derived in the quantum limit with the q-TIP4P/F potential
model.}
\label{fig: exp_qha}
\end{figure}

The experimental phase diagram of ice at temperatures in the range
{[}150, 300{]} K and pressures below 10 GPa is presented in 
Fig.~\ref{fig: exp_qha}a.\cite{dunaeva10}
The broken line shows the boundary between ice an liquid water. Full
lines are the coexistence lines measured for the ice phases. In the
displayed region there appear seven different stable ice phases: Ih,
II, III, V, VI, VII, and VIII. Note that ice XV, the H-ordered counterpart
of ice VI, appears at temperatures lower than 130 K and is not shown
in the figure.\cite{salzmann09}

The QH result derived with the q-TIP4P/F model in the quantum limit
is presented in Fig. \ref{fig: exp_qha}b. The free energy was calculated
for the ice phases listed in Table~\ref{tab:all_ices} by using Eqs.
(\ref{eq: G}) and (\ref{eq:f_v_t}). There are striking differences
between the calculated phase diagram and the experimental data. The
main deviations of the model are:
\begin{itemize}
\item the most stable phase at low pressures is ice Ic instead of ices Ih
or XI,
\item the H-ordered ice IX is stable in the region where experimentally
appears the H-ordered ice II,
\item the H-ordered ice XIV occupies part of the stability region of the
H-disordered ice VI and H-ordered ice XV,
\item the H-disordered ice XII occupies the stability region of ice V.
\end{itemize}
We stress that ices II and V are not stable phases in the calculated
phase diagram. Instead the ice polymorphs IX, XIV, and XII occupy
large regions of stability. Such behavior has not been reported in
any TI simulations of the phase diagram of ice using models based
on the TIP4P potential. \cite{abascal05,vega09,habershon11,mcbride12} 

One may wonder if these unexpected findings are a pathology of the
QHA. Against this point of view it can be argued that the QH phase
diagram of ices Ih, II, and III studied in Ref. \onlinecite{ramirez12b}
is in reasonable agreement with TI simulations. Deviations found between
QH and TI methods for several models (rigid TIP4P/2005 and TIP4PQ/2005,
as well as flexible q-TIP4P/F) were more likely due to structural
differences in the supercell employed for ice III than to limitations
of the QHA.\cite{ramirez12b} For this reason, we consider plausible
that the QHA is providing valid information about the potential model
for ($T,P$) regions and ice phases that have not been previously
studied by TI methods. Thus, the understanding of these unexpected
findings is worth the effort. 

In addition to the information displayed in the phase diagram of Fig.
\ref{fig: exp_qha}b, it is interesting to know the free energy differences
between stable and metastable phases in several regions of the phase
diagram. Free energy differences lower than the threshold of 0.05
kJ/mol are considered within the numerical error of the method and
therefore will not be given a large physical significance. In the
following Section we present a closer look into the calculated free
energies at state points where the most stable phase is either ice
Ic, IX, XIV, or XII.

\begin{table}
\vspace{-3mm}
\caption{QH Gibbs free energy ($G_{0}$) calculated at $T=0$ K and
$P=0$ with the q-TIP4P/F model in the quantum limit. The partition of
$G_{0}$ into lattice ($U_{S,0}$) and zero-point ($U_{Z,0}$) energy is 
given.  $V_{0}$ is the equilibrium volume in
$\textrm{\AA\textthreesuperior}$/molec.
The data for ice III and V correspond to full H-disorder. Energy
units in kJ/mol.}
\label{tab:g_0}
\vspace{3mm}
\begin{tabular}{ccccc}
\hline
q-TIP4P/F & $\quad G_{0\quad}$ & $\quad U_{S,0\quad}$ & $\quad
U_{Z,0\quad}$ & $\quad V_{0\quad}$\tabularnewline
\hline
Ic & 6.97 & -61.77 & 68.74 & 32.35\tabularnewline
XI($Pna2_{1}$) & 6.99 & -61.77 & 68.76 & 32.19\tabularnewline
XI($Cmc2_{1}$) & 7.01 & -61.73 & 68.73 & 32.26\tabularnewline
Ih & 7.01 & -61.74 & 68.75 & 32.23\tabularnewline
IX & 7.22 & -61.40 & 68.62 & 25.60\tabularnewline
II & 7.47 & -60.60 & 68.08 & 25.11\tabularnewline
III (full) & 7.82 & -60.89 & 68.71 & 25.90\tabularnewline
XIV & 7.90 & -60.39 & 68.29 & 22.78\tabularnewline
XIII & 8.18 & -59.97 & 68.15 & 24.01\tabularnewline
V (full) & 8.28 & -60.07 & 68.35 & 23.72\tabularnewline
XII & 8.35 & -59.84 & 68.20 & 22.85\tabularnewline
IV & 8.60 & -59.56 & 68.17  & 22.93\tabularnewline
VI & 8.71 & -59.35 & 68.06 & 22.25\tabularnewline
XV & 8.74 & -59.21 & 67.95 & 22.34\tabularnewline
VIII & 14.05 & -52.81 & 66.86 & 20.55\tabularnewline
VII & 14.15 & -52.80 & 66.95 & 20.43\tabularnewline
\hline
\end{tabular}
\end{table}

\section{Quasi-harmonic free energies\label{sec:QH-free-energies} }

\subsection{\textit{T}=0 K and \textit{P}=0\label{sub:T=00003DP=00003D0}}

The QH Gibbs free energy, $G_{0}$, of an ice phase at $T=0$ K and
$P=0$ is the sum of two energy contributions 

\begin{equation}
G_{0}=U_{S,0}+U_{Z,0}\;.\label{eq:g_t_0_limit}
\end{equation}
$U_{S,0}$ is the static lattice energy for the equilibrium volume,
$V_{0}$, that includes the averaging term for the proton disorder,

\begin{equation}
U_{S,0}=U_{S}(V_{0})+\triangle U_{ave}\;.
\end{equation}
 $U_{Z,0}$ is the zero-point energy calculated as 

\begin{equation}
U_{Z,0}=\sum_{k}\frac{\hbar\omega_{k}(V_{0})}{2}\;.\label{eq:U_z0}
\end{equation}
In Table~\ref{tab:g_0} we collect the values of $U_{S,0}$, $U_{Z,0}$,
and $V_{0}$ of the ice phases studied with the q-TIP4P/F model.

The most stable (i.e. lowest $G_{0})$ phases are ice Ic, XI, and
Ih. Although the predicted order of increasing stability is: Ic>XI($Pna2_{1}$)>XI($Cmc2_{1}$)>Ih,
the free energy differences between them are lower than the threshold
of 0.05 kJ/mol. Such small differences are also conserved at higher
pressures and temperatures. Therefore, the stability of ice Ih, Ic,
and XI is nearly identical for the employed model. For the rest of
the paper we refer to ice Ih as representative for these ice phases
with almost equal free energy.

Then, when compared with other ice polymorphs, ice Ih displays several
distinct properties at $T=0$ K and $P=0$. It has the largest volume
($V_{0}$) per water molecule, the lowest lattice energy ($U_{S,0}$)
and the highest zero-point energy ($U_{Z,0}$) of all ice phases.
The leading factor for the stability of ice Ih at $T=0$ K and $P=$0
is its lowest lattice energy.

It is interesting to note that the nuclear quantum effect causes an
expansion of the equilibrium volume of ice at low temperatures. This
anharmonic effect in the structure of ice is predicted by the QHA
due to zero-point contribution to the free energy in 
Eq.~(\ref{eq:g_t_0_limit}).
This energy term is absent in the classical limit, where the free
energy is equal to the static lattice energy at $T=0$ K. Therefore,
the equilibrium volume, associated to the minimum of the Gibbs free
energy, is different in the quantum and classical limits. We find
that the quantum volume, $V_{0}$, of the ice phases at $T=0$ K (see
Table~\ref{tab:g_0}) are typically 4\% larger than those ones derived
in the classical limit (see $V_{ref}$ values in Table~\ref{tab:all_ices}).

\subsection{Increasing pressure at \textit{T}=0 K\label{sub:Increasing-pressure-at}}

\begin{figure}
\vspace{-1.8cm}
\includegraphics[width= 9cm]{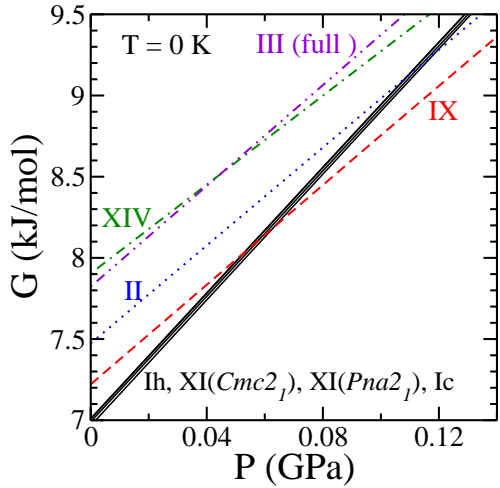}
\vspace{-0.9cm}
\caption{Gibbs free energy, $G,$ of the ice phases with lowest $G$ at
$T=0$
K and pressures below 0.14 GPa. The results correspond to the QHA
using the q-TIP4P/F model. The curves for ices Ih, XI, and Ic are
nearly identical at the displayed energy scale. A phase transition
from ice Ih to ice IX is predicted at $P\sim0.06$ GPa.}
\label{fig:gf1}
\end{figure}

The values of $G_{0}$ and $V_{0}$ in Table~\ref{tab:g_0} allow us
to rationalize the changes in the stability of the ice phases upon
an increase of the pressure at $T=0$ K. A positive pressure will
add a $PV$ term to the free energy $G_{0}$. Obviously the larger
the ice volume the larger the increase in the free energy. Then ice
Ih (with the largest volume) will be destabilized with respect to
all other ice phases upon an increase of pressure. Ice IX is the best
candidate to become stable. It has the lowest value of $G_{0}$ after
that one of ice Ih, and its equilibrium volume is significantly lower
(20\%) than that of ice Ih. 

In Fig. \ref{fig:gf1} we have represented the pressure dependence
up to 0.14 GPa of the Gibbs free energy of the ice phases with lowest
$G$ at $T=0$ K . We have plotted the free energies of ices Ic, Ih,
and XI to show that their small free energy differences at $P=0$
are conserved as pressures increases. We observe that at low pressures
the phase having the minimum free energy is ice Ih. However as the
pressure increases above $0.06$ GPa, ice IX becomes more stable than
ice Ih.

\begin{figure}
\vspace{-1.8cm}
\includegraphics[width= 9cm]{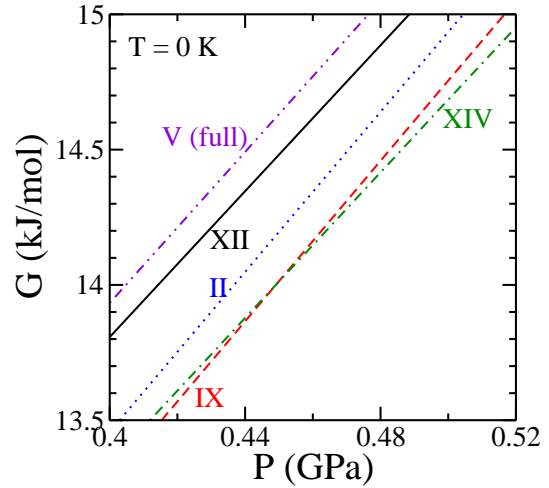}
\vspace{-0.9cm}
\caption{Gibbs free energy, $G,$ of the ice phase with lowest $G$ at
$T=0$
K and pressures in the interval {[}0.4,0.52{]} GPa. The results are
derived by the QHA using the q-TIP4P/F model. A phase transition from
ice IX to ice XIV is predicted at $P\sim0.45$ GPa.}
\label{fig:gf2}
\end{figure}

A further increase of the pressure will stabilize another ice phase
with even lower volume than ice IX. In Fig. \ref{fig:gf2} we show
the Gibbs free energy, $G$, of several ice phases in the range $0.42<P<0.48$
GPa. The crossing of the free energy lines of ice IX and XIV at 0.45
GPa is the fingerprint for a phase transition from ice IX to ice XIV. 

Given a pressure $P$, for thermodynamic consistency in the low-temperature
limit ($T\to0$ K) one expects the stable ice polymorph to be an H-ordered
phase. This does not appear to be always the case in our calculations.
Thus, at $P<60$ MPa we find cubic ice Ic to be the low $T$ stable
phase. However, as indicated above, free-energy differences between
ice Ic and the H-ordered ice XI are smaller than our sensitivity limit
($\sim0.05$ kJ/mol). Also, ice VI appears to be the stable phase
at low temperatures in the region between 3 and 5 GPa. In this case
one would expect the corresponding H-ordered phase (ice XV) to be
the stable polymorph, but its free energy for $T\to0$ K is higher
than that of ice VI. This problem may be related to the possibility
that the q-TIP4P/F potential could favor an H-ordered pattern different
from that corresponding to ice XV, as those discussed earlier in the
literature.\cite{salzmann09}

\subsection{Stability of ice IX versus ice II}

A consequence of the stability of ice IX is that ice II does not appear
(i.e., it is metastable) in the QH phase diagram of Fig. \ref{fig: exp_qha}b.
This metastability of ice II disagrees obviously with the experimental
phase diagram. In Figs. \ref{fig:gf1} and \ref{fig:gf2} the free
energy $G$ of ice II is larger than that of ice IX. The difference
is nearly independent of the pressure, as the $G(P)$ curves of ices
IX and II are approximately parallel. If we measure the free energy
difference between these phases at $T=0$ by the ordinates in the
origin of Fig. \ref{fig:gf1} (i.e., the $G_{0}$ values of ice IX
and II in Table~\ref{tab:g_0}), one gets

\begin{equation}
\Delta G_{0}(\textrm{II-IX})=0.25\;\textrm{kJ/mol}\;.
\end{equation}
The positive value implies that ice IX is more stable than ice II.
The energy partitioning of $G_{0}$ in Table~\ref{tab:g_0} shows that
the leading term for the larger stability of ice IX is its lower lattice
energy.

\begin{table}
\vspace{-3mm}
\caption{Gibbs free energy ($G_{0,cla}$ in kJ/mol) calculated with
the rigid
TIP4P/2005 model for several ice phases at $T=0$ K and $P=0$ in
the classical limit. $V_{0,cla}$ is the equilibrium volume in
$\textrm{\AA\textthreesuperior}$/molec. }
\label{tab:TIP4P_2005}
\vspace{3mm}
\begin{tabular}{ccc}
\hline
TIP4P/2005 & $\quad G_{0,cla\quad}$ & $\quad
V_{0,cla}\quad$\tabularnewline
\hline
Ih & -62.99 & 31.34\tabularnewline
IX & -62.71 & 24.87\tabularnewline
II & -62.13 & 24.30\tabularnewline
XIV & -61.72 & 22.10\tabularnewline
\hline
\end{tabular}
\end{table}

Note that if the $G(P)$ curve of ice IX were omitted from Fig.~\ref{fig:gf1},
then the first transition as a function of pressure would correspond
to the crossing of the Ih and II free energy curves at 0.11 GPa. This
Ih-II transition was shown in the study of the coexistence of ices
Ih, II, and III in Fig. \ref{fig: I_II_III}. 

Given the large stability of ice IX predicted by the q-TIP4P/F model,
we want to address the following question: Is this stability a consequence
of the flexibility of the model or it has its origin in the TIP4P-character
of the potential? 

To this aim we have calculated the QH free energies of several ice
phases by using the rigid TIP4P/2005 model. This model was parameterized
for water simulations in the classical limit. In this limit, at $T=0$
K and $P=0$, the Gibbs free energy and equilibrium volume of an ice
phase are simply given by

\begin{equation}
G_{0,cla}\equiv\overline{U}_{S,ref}\;,\label{eq:G_0_cla}
\end{equation}

\begin{equation}
V_{0,cla}\equiv V_{ref}\;.
\end{equation}
 For reference, the values of $G_{0,cla}$ and $V_{0,cla}$ calculated
with the TIP4P/2005 model for ices Ih, XI, II, and XIV are shown in
Table~\ref{tab:TIP4P_2005}. Note that in this case

\begin{equation}
\Delta G_{0,cla}(\textrm{II-IX})=0.58\;\textrm{kJ/mol}\;.
\end{equation}
This free energy difference between ice IX and II is even larger than
that found for the flexible q-TIP4P/F model.

The classical QH phase diagram of ice Ih, II, and IX was calculated
with the rigid TIP4P/2005 model at temperatures up to 300 K and pressures
below 0.6 GPa. We find that ice IX is more stable than ice II in the
whole studied ($T,P$) region. Therefore ice II is metastable in the
classical phase diagram of the TIP4P/2005 model. This result is identical
to that found for the flexible q-TIP4P/F model. Our conclusion is
that the large stability of ice IX is a property of the TIP4P-character
of the model, and not a consequence of the added molecular flexibility. 

We believe that limitations inherent to the exploration of the phase
diagram by TI methods is the reason why the stability of ice IX and
the metastability of ice II has not been detected in previous studies
using TIP4P-like models.\cite{vega09,mcbride12} An advantage of the
QHA is that the brute force calculation of free energies allows a
thorough exploration of state points for all ice phases.

\subsection{Increasing pressure at \textit{T}=250 K\label{sub:QHA-free-energies-250K}}

\begin{figure}
\vspace{-1.8cm}
\includegraphics[width= 9cm]{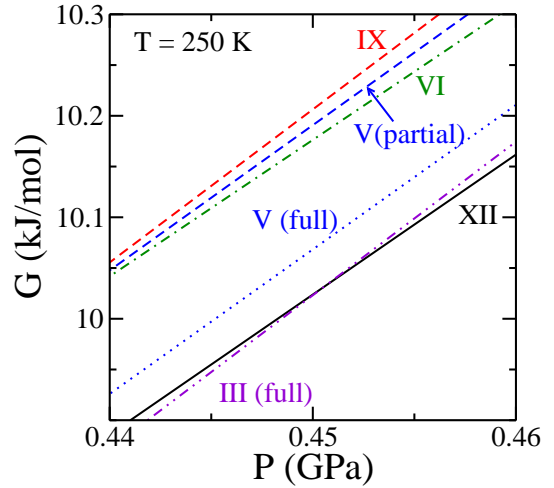}
\vspace{-0.9cm}
\caption{Gibbs free energy, $G,$ of the ice phase with lowest $G$ at
$T=250$ K and pressures in the interval {[}0.44,0.46{]} GPa. The results
correspond to the q-TIP4P/F model and the QHA. A phase transition from 
ice III to ice XII appears at $P\sim0.45$ GPa.}
\label{fig:gft2}
\end{figure}

An unexpected result of the calculated phase diagram at temperatures
around 250 K is that ice XII, the H-disordered counterpart of ice
XIV, occupies the stability region where one would expect to find
ice V as stable phase (see Fig. \ref{fig: exp_qha}). The pressure
dependence of the free energy of the ice phases with lowest $G$ is
presented in Fig. \ref{fig:gft2} at 250 K. The crossing of the $G(P)$
curves of ice III and ice XII at $P=0.45$ GPa indicates that ice
XII becomes the stable phase at 250 K for pressures higher than 0.45
GPa. 

It is interesting to note that at the pressures shown in Fig. \ref{fig:gft2}
the free energy of ice V (with full H-disorder) is only slightly higher
($\sim0.06$ kJ/mol) than that of ice XII. Besides, ice XII displays
lower volume than ice V. Therefore an increase in the pressure will
always stabilize ice XII with respect to ice V. At $T=250$ K the
QHA predicts that the coexistence pressure for ices V-XII is 0.33
GPa. The equilibrium volumes at this state point ($T=250$ K, $P=0.33$
GPa) are 23.1 $\lyxmathsym{\AA}\text{\textthreesuperior}$/molec.
and 23.8 $\lyxmathsym{\AA}\text{\textthreesuperior}$/molec., for
ice XII and ice V, respectively. 

Note that at 250 K the pressure interval where ice V appears as stable
phase in the experimental phase diagram is about {[}0.35,0.6{]} GPa
(see Fig. \ref{fig: exp_qha}a). In this pressure range the q-TIP4P/F
model predicts that the free energy difference between ices V and
XII increases from a vanishingly small value (at 0.35 GPa) to a maximum
value of 0.1 kJ/mol (at 0.6 GPa). Therefore free energy differences
between ice XII and V are relatively small. Similar values for the
free energy of ice V and XII have been already reported for the TIP4P/2005
model at $P=0.5$ GPa  by TI simulations in the classical limit.\cite{vega09}

\section{Quantum vs. classical phase diagram\label{sec:Quantum-vs.-classical}}

\begin{figure}
\vspace{-1.8cm}
\includegraphics[width= 9cm]{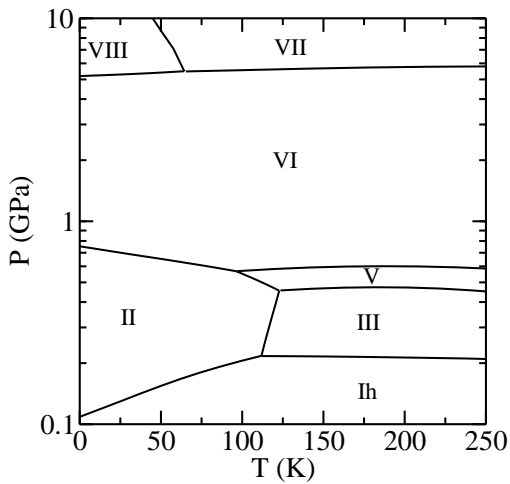}
\vspace{-0.9cm}
\caption{QHA of the phase diagram of ice using the q-TIP4P/F model in
the quantum
limit. Ices III and V are full H-disordered. The following ice phases
were omitted from the QH calculation: ice IX, XIV, and XII. }
\label{fig:all}
\end{figure}

The QH phase diagram of ice shown in Fig. \ref{fig: exp_qha}b for
the q-TIP4P/F model is characterized by the stability of ice IX, XII,
and XIV over large regions of the ($T,P$) plane. However, these phases
have not been reported as stable ones in previous studies. For this
reason our calculated phase diagram has little resemblance to previous
ones derived by using rigid TIP4P-like potentials in combination with
TI methods.\cite{sanz04,vega09,mcbride12}

Then, it is interesting to recalculate our phase diagram of ice under
omission of the phases IX, XII, and XIV. In addition we will consider
ice Ih as unique representative of the phases Ih, Ic, and XI. 

The quantum limit of the new phase diagram of the q-TIP4P/F model
is presented in Fig. \ref{fig:all}. Now we find the following stable
phases: Ih, II, III, V, VI. VII and VIII. Both ices III and V correspond
to full H-disordered lattices. This phase diagram is in reasonable
qualitative agreement to the experimental one. Moreover, it is also
in reasonable agreement to those phase diagrams derived for TIP4P-like
potentials.\cite{sanz04,vega09,mcbride12} 

Concerning the thermodynamic consistency of these results for $T\to0$
K, we emphasize that the H-disordered ices Ih and VI cannot strictly
be the low-temperature stable phases for any pressure $P$. For this
question, the arguments are the same as those given above in Sec.
\ref{sub:Increasing-pressure-at} and are not repeated here.

Our results in Fig. \ref{fig:all} include the averaging of the lattice
energy over H-disorder for ice III, V, and VI. None of the previously
published phase diagrams calculated with TIP4P-like potentials include
any kind of disorder averaging. In fact, different single H-isomers
of ice III seem to have been employed for the calculations with several
TIP4P-like potentials (i.e., TIP4P,\cite{sanz04} TIP4P/2005,\cite{vega09}
TIP4PQ/2005\cite{mcbride12}). If the ice III structure employed for
each potential model has a different H-configuration, then the lattice
energy of ice III may be affected by an uncontrolled factor. This
situation makes it difficult to draw definitive conclusions about
differences found in calculated phase diagrams with different H-isomers
of ice III. This uncertainty should affect not only ice III, but also
the stability region of other phases (Ih, II, V) having a boundary
with ice III.

\begin{figure}
\vspace{-1.8cm}
\includegraphics[width= 9cm]{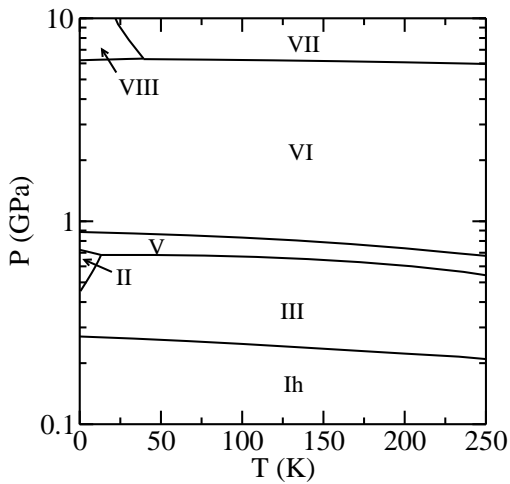}
\vspace{-0.9cm}
\caption{The phase diagram of Fig.~\ref{fig:all} is calculated with the
q-TIP4P/F model  in the classical limit.}
\label{fig:all_c}
\end{figure}

It is interesting to compare the quantum phase diagram in 
Fig.~\ref{fig:all} with that one calculated within the classical limit
(Fig.~\ref{fig:all_c}). The main difference
between both limits is related to the stability region of ice II.
It is much larger in the quantum case. The origin of this quantum
effect is related to the lower zero-point energy of ice II, in comparison
to ices Ih, III, and V (see the $U_{Z,0}$ values given in 
Table~\ref{tab:g_0}).
The quantum stabilization of ice II has been already discussed in
Ref. \onlinecite{ramirez12b}.

An additional difference between quantum and classical phase diagrams
is that the coexistence lines that are nearly horizontal (i.e., Ih-III,
III-V, V-VI) appear shifted to higher pressures in the classical limit.
This effect has been also reported in TI simulations with the rigid
TIP4PQ/2005 model.\cite{mcbride12} The rising of the pressure occurs
at all studied temperatures. This quantum effect is most easily explained
at $T=0$ K. Let us consider ices VI and V as an example. In the quantum
limit, the free energy difference between ice VI and V at $T=0$ K
and $P=0$ is (see the $G_{0}$ data in Table~\ref{tab:g_0}),

\begin{equation}
\triangle G_{0}(\mathrm{VI-V})=0.4\;\mathrm{kJ/mol}\;.\label{eq:g_dif}
\end{equation}
However, in the classical limit one gets using Eq. (\ref{eq:G_0_cla})
and the values of Table~\ref{tab:all_ices},

\begin{equation}
\triangle G_{0,cla}(\mathrm{VI-V})=0.7\;\mathrm{kJ/mol}\;.\label{eq:g_dif_cla}
\end{equation}
In both cases $\Delta G_{0}>0,$ i.e., ice VI is less stable than
ice V. However, in the quantum case $\Delta G_{0}$ is lower. The
reason is that the zero-point energy (see the $U_{Z,0}$ data in 
Table~\ref{tab:g_0}) tends to stabilize ice VI with respect to ice V 
by $\sim-0.3$ kJ/mol. 

On the other side, although the volume of ice VI is lower than that
of ice V (see $V_{0}$ data in Table~\ref{tab:g_0}), the volume differences
are found to be nearly the same in the quantum 
($\Delta V_{0}=-1.5\quad\textrm{\AA\textthreesuperior}$/molec.)
and classical cases 
($\Delta V_{0,cla}=-1.4\quad\textrm{\AA\textthreesuperior}$/molec.,
see the $V_{ref}$ data in Table~\ref{tab:all_ices}). 

By increasing the pressure, the phase with lower volume (ice VI) will
become more stable. The pressure needed to stabilize ice VI with respect
to ice V is roughly given by

\begin{equation}
P\thicksim-\frac{\Delta G_{0}}{\Delta V_{0}}\;.\label{eq:p}
\end{equation}
Note that while the denominator is similar in both quantum and classical
limits, the numerator is lower in the quantum case. Therefore the
coexistence pressure for ices V-VI at $T=0$ K is reduced in the quantum
limit with respect to the classical one.

A similar argument explains why the transitions Ih-III and III-V are
displaced to higher pressures in the classical case.\cite{mcbride12}

\section{Conclusions\label{sec:Conclusions}}

The phase diagram of ice has been studied by a quasi-harmonic approximation
using the flexible q-TIP4P/F model of water. The simplicity of this
approach allows us to include all known ice polymorphs (except ice
X) and all state points for $T<$ 400 K and $P<$ 10 GPa. 

Surprisingly the simple QHA seems to be accurate enough to reproduce
free energy differences between ice phases, in spite of the large
complexity and variety in their crystal structures. This conclusion
about the success of the QHA is derived using a simple model potential,
but its validity is expected to be largely independent of the model.
Therefore it opens a route for the study of the whole phase diagram
of ice by ab initio electronic structure methods.

The H-disorder of many ice phases is an additional complication in
the calculation of their phase diagram. The QHA has allowed us to
quantify the influence of this effect. The disorder averaging of the
lattice energy of ice III has been proven to be important to obtain
a converged phase diagram, at least using TIP4P-like models and ice
III supercells with full H-disorder. Disorder averaging of vibrational
free energies has been shown to be comparatively less important. In
addition to ice III, the disorder averaging of the lattice energy
of ice XII and V has been shown to be also significant for the ice
stability. 

We stress that phase diagrams calculated using a single random H-isomer
of ice III may be affected by an uncontrolled energetic factor that
can be highly significant in the final result. An immediate consequence
of this is that comparison of phase diagrams calculated using a single,
but different, H-isomer of ice III might not be physically sound.
The reason is that the stability of the disordered phase may depend
strongly on the employed H-isomer.

The QH phase diagram of ice with the flexible q-TIP4P/F model has
been calculated by performing a disorder averaging of the lattice
energy of the H-disordered ice phases. We have found an unexpected
large stability of several ice phases, specially the H-ordered ices
IX and XIV, and also the H-disordered ice XII. The presence of these
phases in the calculated phase diagram implies that both ice II and
V are metastable phases with the q-TIP4P/F model. This finding disagrees
with the experimental phase diagram. We have checked that ice IX remains
more stable than ice II if the phase diagram is calculated using the
rigid TIP4P/2005 model in the classical limit. Our conclusion is that
the larger stability of ice IX with respect to ice II is a property
related to the TIP4P-character of the model and not to the explicit
treatment of the molecular flexibility. 

The QH free energy and volume of several ice phases have been analyzed
at $T=0$ K and $P=0$. The free energy has been partitioned into
lattice and zero-point energies. These contributions are important
magnitudes in the analysis of the stability of the ice phases as a
function of pressure.

By excluding ices IX, XIV, and XII from the calculation, we find that
the phase diagram of the q-TIP4P/F model shows qualitative agreement
to both experimental and previously simulated ones by using TIP4P-like
models. The comparison of the quantum and classical limits shows several
differences. The most important are the increase in the stability
of ice II and the shift of the coexistence lines III-V and V-VI to
lower pressures in the quantum case. Similar conclusions were reached
previously in Ref. \onlinecite{mcbride12}. Differences in the zero-point
energies of the ice phases provide an explanation for these effects.

\acknowledgments

This work was supported by MEC (Spain) through Grant No. FIS2012-31713,
and by Comunidad Autónoma de Madrid through project MODELICO-CM/S2009ESP-1691.

\bibliographystyle{apsrev}

\end{document}